\documentclass[aps,showpacs,preprintnumbers,amsmath,amssymb]{revtex4}
\UseRawInputEncoding
\usepackage{dcolumn}
\usepackage[dvips]{epsfig}

\usepackage{float}
\usepackage{graphics,epsfig}
\usepackage{graphicx}
\usepackage{multirow}
\usepackage{dcolumn}
\usepackage{bm}
\usepackage{gensymb}

\begin{document}
\baselineskip=0.8 cm
\title{Tidal effects in 4D Einstein-Gauss-Bonnet Black Hole Spacetime}

\author{Jing Li$^{1}$
Songbai Chen$^{1,2}$\footnote{Corresponding author: csb3752@hunnu.edu.cn},
Jiliang Jing$^{1,2}$ \footnote{jljing@hunnu.edu.cn}}
\affiliation{ $ ^1$ Department of Physics, Key Laboratory of Low Dimensional Quantum Structures
and Quantum Control of Ministry of Education, Synergetic Innovation Center for Quantum Effects and Applications, Hunan
Normal University,  Changsha, Hunan 410081, People's Republic of China
\\
$ ^2$Center for Gravitation and Cosmology, College of Physical Science and Technology, Yangzhou University, Yangzhou 225009, People's Republic of China}

\begin{abstract}
\baselineskip=0.6 cm
\begin{center}
{\bf Abstract}
\end{center}

 We have investigated tidal forces and geodesic deviation motion in the 4D-Einstein-Gauss-Bonnet spacetime. Our results show that tidal force and geodesic deviation motion depend sharply on the sign of Gauss-Bonnet coupling constant. Comparing with Schwarzschild spacetime, the strength of tidal force becomes stronger for the negative Gauss-Bonnet coupling constant, but is weaker for the positive one. Moreover,  tidal force behaves like those in the Schwarzschild spacetime as the coupling constant is negative, and like those in Reissner-Nordstr\"{o}m black hole as the constant is positive. We also present the change of geodesic deviation vector with Gauss-Bonnet coupling constant under two kinds of initial conditions.

\end{abstract}

 \maketitle
\newpage
\section{Introduction}

A broad area of alternative theories of gravity have been proposed to add extra higher curvature corrections to Einstein's general relativity.  Gauss-Bonnet gravity is such a higher correction theory where the curvature correction is precisely given by the so-called Gauss-Bonnet term, which is a combination of quadratic curvature terms.
Generally, the Gauss-Bonnet term in the four dimensional spacetime is a total
derivative and then it does not contribute to the equations
of motion. However, through multiplying the higher dimensional
Gauss-Bonnet term by the factor $1/(D-4)$,  Glavan \textit{et al} \cite{4D-EGB} recently find that the original disappeared Gauss-Bonnet contributions in the four dimensional spacetimes now present finite nontrivial effects due to the interaction with the divergent factor $1/(D-4)$. By solving the corresponding equation of motion, they obtain some new spherically symmetric black hole solutions  in the 4D Einstein-Gauss-Bonnet Gravity. The physical properties and the novel observable effects for these black hole solutions have been explored in refs.\cite{4D-EGB2,shadow1,shadow2,GGL1,GGL2,GGL3,GGL4,GGL5}. However, there are also some criticisms on this theory of Gravity \cite{dispute1,dispute2,dispute3,dispute4,dispute5}. Especially, the original suggestion in \cite{4D-EGB} is found to be in contradiction with the Lovelock theory, and the regularization scheme  \cite{4D-EGB} does not produce the four dimensional theory of gravity \cite{pure4D}. Another ``regularized"
version of the 4D Einstein-Gauss-Bonnet gravity proposed in \cite{rega1,rega2}, which can be described effectively by a particular subclass of the Horndeski theory,  contains an additional dynamical scalar field but lacks the quadratic kinetic term and thus gives rise to the infinite strong coupling problem \cite{scalar-tensor,Amplitudes}.  Recently, with the Arnowitt-Deser-Misner (ADM) decomposition, a well-defined theory is proposed to serve as a consistent realization of the 4D Einstein-Gauss-Bonnet gravity with two dynamical degrees of freedom but breaking the temporal diffeomorphism invariance \cite{consistent theory}. Luckily, the black hole solution obtained by the naive regularization \cite{4D-EGB} also is found to be an exact solution in this full theory \cite{consistent theory}. Therefore, it is very safe to consider test fields in the background of these black holes using the same regularization scheme. Along this spirit, the Einstein-Gauss-Bonnet gravity in the 4D spacetime has been investigated extensively on various aspects \cite{4D-EGB3,4D-EGB4,4D-EGB5,4D-EGB6,4D-EGB7,Holographic,Hawking,Cosmology,EDGR1,EDGR2,EDGR3,EDGR4,EDGR5,EDGR6,EDGR7,EDGR8,
EDGR9,EDGR9a,EDGR10,EDGR11,EDGR12,EDGR13,EDGR14,EDGR15,EDGR16,EDGR17,EDGR18}.

Tidal phenomena are common in our Universe and tidal forces play an important role in astrophysical context such as tidal disruption events. It is well known that tidal force is caused by a difference in the strength of gravity between two points, which yields shape deformation of a body in gravitational field.  In a Schwarzschild spacetime, a body falling towards the event horizon gets stretched in the radial direction and compressed in the angular one \cite{s1,s2,s3,Schwarzschild TF}. In Reissner-Nordstr\"{o}m spacetime \cite{RNBH,RNBH1}, it is found that the tidal effect depends on the
charge-to-mass ratio of the black hole and the position of the body. Moreover, the radial or angular component of the tidal force changes its sign as the body falls towards to the center of black hole. The tidal forces have been investigated
in some other spherically symmetric spacetimes including regular black holes \cite{CH,RegularBH}, Kiselev black holes \cite{Kiselev}, and naked singularity \cite{naked singularity}. The study also shows \cite{TDE0} that a star in ergosphere of a rotating black hole can be broken up due to tidal interaction and then emits subsequently a jet composed of the debris, which could provide a potential mechanism for the jets production in black hole spacetime. Subsequently, the investigation of tidal effects have also been performed in various black holes \cite{TDE,TDE1,TDE2,TDE3,TDE4,kerr}. The main purpose of this paper is to study the tidal effects in the 4D-Einstein-Gauss-Bonnet black hole spacetime and to probe how the Gauss-Bonnet coupling constant affect tidal forces and the motion of geodesic deviation vector in the four dimensional spacetime.

The paper is organized as follows: In Sec.II, we briefly review the 4D-Einstein-Gauss-Bonnet black hole and the geodesics equation. In Sec.III, we investigate tidal forces on a body falling free along radial direction in the 4D Einstein-Gauss-Bonnet black hole spacetime. Finally, we end the paper with a summary.

\section{Geodesics in 4D-Einstein-Gauss-Bonnet Black Hole spacetime}
Let us now to review briefly the 4D static Einstein-Gauss-Bonnet black hole solution obtained in \cite{4D-EGB}.
The usual Einstein-Gauss-Bonnet gravity action in $D$ dimensional spacetime can be expressed as
\begin{equation}\label{Metric}
S=-\frac{1}{16\pi G}\int d^{D}x \sqrt{-g}\bigg[R+\alpha(R_{\mu\nu\rho\sigma}R^{\mu\nu\rho\sigma}-4R_{\mu\nu}R^{\mu\nu}+R^2)\bigg].
\end{equation}
where $\alpha$ is Gauss-Bonnet coupling constant. Rescaling the coupling constant $\alpha\rightarrow \frac{\alpha}{D-4}$ and taking the limit $D\rightarrow 4$, Glavan \textit{et al}  obtained firstly a neutral 4D static Einstein-Gauss-Bonnet black hole solution by solving the corresponding equation of gravity field \cite{4D-EGB}.  The metric of the black hole has a form
\begin{equation}\label{Metric}
d s^{2}=-f(r) d t^{2}+\frac{d r^{2}}{f(r)}+r^{2} d \theta^{2}+r^{2} \sin ^{2} \theta d \phi^{2},
\end{equation}
with
\begin{eqnarray}\label{gtt}
f(r)=1+\frac{r^{2}}{2 \alpha}\left(1\pm\sqrt{1+\frac{8 \alpha M}{r^{3}}}\right),
\end{eqnarray}
where $M$ is black hole mass. Here we focus only on the ``minus" branch of above solution because it can asymptotically go over to the Schwarzschild black hole with the correct mass sign. It is easy to find that the black hole horizons satisfy equation
 $f(r)=0$ and its roots are
 \begin{equation}\label{Horizon}
r_{\pm}=M \pm \sqrt{M^2-\alpha},
\end{equation}
where $r_{+}$ and $r_{-}$ correspond to the event horizon radius and the Cauchy horizon one, respectively. It is obvious that the event horizon radius decreases with the constant $\alpha$, while the Cauchy horizon radius increases, which is also shown in Fig.\ref{as0}. As $\alpha=M^{2}$, the event horizon overlaps with the Cauchy horizon and then the black hole becomes an extreme black hole. As $\alpha>M^{2}$, the horizons disappear and then it becomes a naked singularity. As $\alpha<0$, one can find that $r_-$ is negative and then there is a single horizon for the black hole \eqref{Metric} as in the usual Schwarzschild black hole case.

In the 4D-Einstein-Gauss-Bonnet black hole spacetime \eqref{Metric}, a massive particle moving along the radial direction satisfies the motion equation \cite{s1}
\begin{equation}\label{radialmotion}
-f(r) \dot{t}^{2}+f^{-1}(r) \dot{r}^{2}=-1,
\end{equation}
where dot represents the differentiation with respect to
the proper time $\tau$. Since the coefficients in metric (\ref{Metric}) is independent of the coordinates $t$ and $\phi$,
there exist two conserved quantities for a massive particle moving along the geodesics, i.e.,
the energy $E$ and angular momentum $L$. Substituting the particle's  energy $E=f(r) \dot{t}$ into Eq.\eqref{radialmotion},
one has
\begin{equation}\label{4Denergy}
\dot{r}^{2}=E^{2}-f(r).
\end{equation}
And then the Newtonian radial acceleration for the particle in the 4D-Einstein-Gauss-Bonnet black hole spacetime (\ref{Metric}) can be expressed as
\begin{equation}\label{Nra}
A^{(R)} \equiv \ddot{r}=-\frac{f^{\prime}(r)}{2}=\frac{M}{r^2}\bigg[\frac{4}{1+\sqrt{1+\frac{8\alpha M}{r^{3}}}}-\frac{3 }{ \sqrt{1+\frac{8 \alpha M}{r^{3}}}}\bigg].
\end{equation}
It increases with the  coupling constant $\alpha$, which means that the Newtonian radial acceleration in the 4D-Einstein-Gauss-Bonnet black hole spacetime is larger than that in the usual Schwarzschild case where $A^{(R)}=-\frac{M}{r^{2}}$.
From Eq.\eqref{4Denergy}, the test particle falling freely
from rest at $r=b>r_+$ would bounce back in its radial motion at the turning point
$r=R^{\text {stop }}$, i.e.,
\begin{equation}\label{Rstop}
R^{\text {stop }}=\frac{4\alpha M}{b^{2}+\sqrt{b(b^{3}+8 \alpha M)}}.
\end{equation}
For the negative $\alpha$, one can find that $R^{\text {stop }}<0$, which means that there is no turning point for the test particle falling freely as in the Schwarzschild case. For the positive $\alpha$, we find that $R^{\text {stop }}$ decreases monotonously with the initial radial coordinate $b$ of particle. Thus, for the test particle falling freely
from rest at $r=b>r_+$, one has
\begin{equation}\label{Rstop}
R^{\text {stop }}<R^{\text {stop }}|_{b=r_+}=r_-,
\end{equation}
which means that the position of the turning point for the above test particle is always located inside the Cauchy horizon (see also in Fig.\ref{as0}) and the behavior of $R^{\text {stop }}$ in 4D-Einstein-Gauss-Bonnet black hole spacetime \eqref{Metric} is similar to those in other static black hole spacetimes.

\section{Tidal Forces on a neutral Body in Radial Free Fall in 4D Einstein-Gauss-Bonnet Black Hole}

\begin{figure}
\begin{center}
  \includegraphics[width=6cm ]{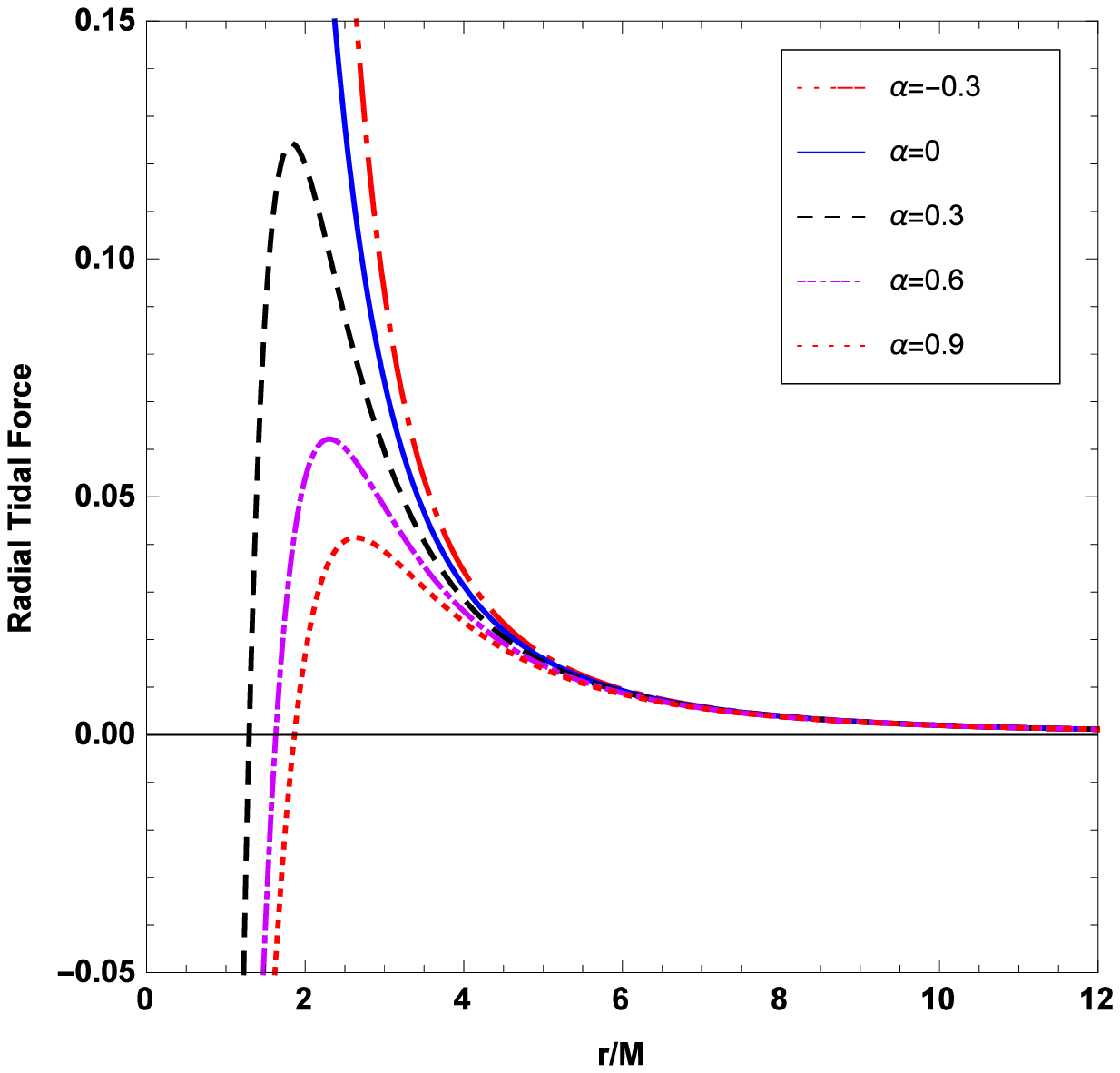}\quad\includegraphics[width=6cm ]{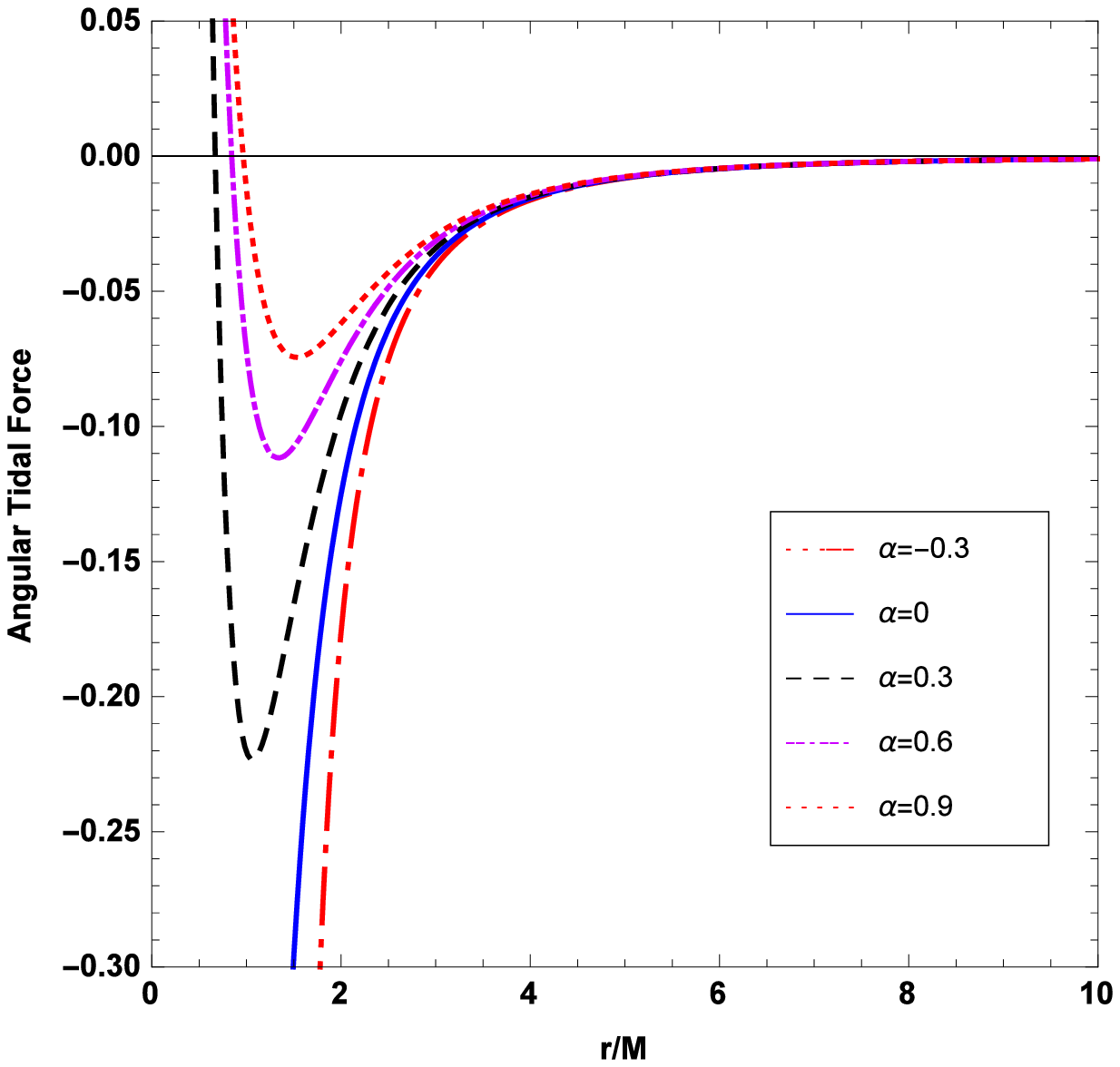}
\caption{The tidal forces in 4D Einstein-Gauss-Bonnet black hole spacetime with different $\alpha$. The position of $r_{+} $ and $r_{-} $ are marked with  the chartreuse and the deep red, respectively. The left panel is for the tidal force and the right panel is for the angular one. Here we set $M=1$ and $b=100M$. }
\label{as1}
\end{center}
\end{figure}

\begin{figure}
\begin{center}
\includegraphics[width=6cm]{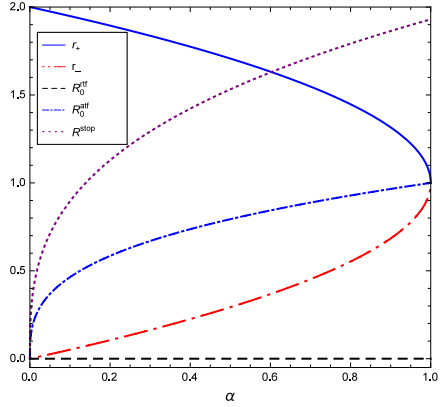}
\caption{ The change of $R^{\text {stop }}, R_{0}^{\mathrm{rtf}}, R_{0}^{\text {atf }}, r_{-}, \text {and } r_{+}$ with the coupling constant $\alpha$.  Here we set $M=1$ and $b=100$.}
\label{as0}
\end{center}
\end{figure}

We are now in position to study the tidal forces on the massive particle in 4D-Einstein-Gauss-Bonnet black hole spacetime \eqref{Metric}. In general, the tidal forces are governed by a geodesic deviation equation of a spacelike vector $\eta^{\mu }$ representing
the distance between two infinitesimally close particles following geodesics \cite{s1,s2,s3,Schwarzschild TF,RNBH,RNBH1}. The geodesic deviation equation can be expressed as
\begin{equation}\label{geodesic deviation}
\frac{D^{2} \eta^{\mu}}{D \tau^{2}}-R_{\sigma \nu   \rho}^{\mu} \nu  ^{\sigma} \nu  ^{\nu  } \eta^{\rho}=0.
\end{equation}
where $\nu^{\mu} $ is the unit vector tangent to the geodesic. For the 4D-Einstein-Gauss-Bonnet black hole spacetime \eqref{Metric}, we can introduce the tetrad basis describing a freely falling frame
\begin{eqnarray}\nonumber
&&\hat{e}_{\hat{0}}^{\mu}=\left(\frac{E}{f},-\sqrt{E^{2}-f}, 0,0\right), \quad\quad\quad
\hat{e}_{\widehat{1}}^{\mu}=\left(\frac{-\sqrt{E^{2}-f}}{f}, E, 0,0\right),\\\nonumber
&&\hat{e}_{\hat{2}}^{\mu}=r^{-1}\left(0,0, 1,0\right),\quad\quad\quad\quad\quad\quad
\hat{e}_{\hat{3}}^{\mu}=(r \sin \theta)^{-1}\left(0,0,0,1\right),\
\end{eqnarray}
which satisfy the orthogonal conditions
\begin{equation}\label{ng}
\hat{e}_{\hat{\alpha}}^{\mu} \hat{e}_{\hat{\beta}\mu}=\eta_{\hat{\alpha} \hat{\beta}}.
\end{equation}
Here the indices with hat are tetrad basis indices, while the indices without hat are coordinate
basis indices.  $\eta_{\hat{\alpha} \hat{\beta}}$ is the Minkowski metric. The geodesic deviation vector $\eta^{\hat{\mu}}$ can be expanded as $\eta^{\hat{\mu}}=\hat{e}_{{\nu}}^{\hat\mu} \eta^{\nu}$ with a fixed temporal component $\eta ^{\hat{0} }=0$ \cite{s1} . Moreover, we have $e_{\hat{0}}^{\mu}=\nu ^{\mu }$ .
As in a usual spherically symmetric black hole spacetime, the non-vanishing independent components of the Riemann curvature tensor in the 4D-Einstein-Gauss-Bonnet black hole spacetime (\ref{Metric}) can be expressed as
\begin{eqnarray}\nonumber
&&R_{010}^{1}=\frac{f f^{\prime \prime}}{2},  \quad\quad\quad\quad\quad\quad    R_{212}^{1}=-\frac{r f^{\prime}}{2}, \\\nonumber
&&R_{313}^{1}=-\frac{r f^{\prime}}{2} \sin ^{2} \theta,   \quad\quad\quad  R_{020}^{2}=\frac{f f^{\prime}}{2 r},\\\nonumber
&&R_{323}^{2}=(1-f) \sin ^{2} \theta, \quad\quad\quad   R_{030}^{3}=\frac{f f^{\prime}}{2 r}.\
\end{eqnarray}
With the relationship $R_{\hat{\beta} \hat{\gamma} \hat{\delta}}^{\hat{\alpha}}=e_{\mu}^{\hat{\alpha}} e_{\hat{\beta}}^{\nu} e_{\hat{\gamma}}^{\rho} e_{\hat{\delta}}^{\sigma} R_{\nu \rho \sigma}^{\mu}$ \cite{s1,s2,s3}, one can get the Riemann curvature tensor in the tetrad basis
\begin{equation}\label{ff-Riemann}
R_{\hat{1} \hat{0} \hat{1}}^{\hat{0}}=-\frac{f^{\prime \prime}(r)}{2}, \quad\quad\quad R_{\hat{2} \hat{0} \hat{2}}^{\hat{0}}=R_{\hat{3} \hat{0}\hat{3}}^{\hat{0}}=R_{\tilde{2} \hat{1} \hat{2}}^{\hat{1}}=R_{\hat{3} \hat{1} \hat{3}}^{\hat{1}}=-\frac{f^{\prime}(r)}{2 r}, \quad\quad\quad R_{\hat{3} \hat{2}\hat{3}}^{\hat{2}}=\frac{1-f(r)}{r^{2}}.
\end{equation}
Using these expressions, one can find that the equations for tidal forces in radial free-fall
reference frames can be expressed as
\begin{equation}\label{RTF}
\ddot{\eta}^{\hat{1}}=-\frac{f^{\prime \prime}}{2} \eta^{\hat{1}}=\frac{4M}{r^3}\bigg[\frac{1}{1+\sqrt{1+\frac{8\alpha M}{r^{3}}}}-\frac{9 \alpha M}{r^3 (\sqrt{1+\frac{8 \alpha M}{r^{3}}})^3}\bigg] \eta^{\hat{1}},
\end{equation}
\begin{equation}\label{ATF}
\ddot{\eta}^{\hat{i}}=-\frac{f^{\prime}}{2 r} \eta^{\hat{i}}=\frac{M}{r^2}\bigg[\frac{4}{1+\sqrt{1+\frac{8\alpha M}{r^{3}}}}-\frac{3 }{ \sqrt{1+\frac{8 \alpha M}{r^{3}}}}\bigg] \eta^{\hat{i}}.
\end{equation}
where $i=2,3$ denote the angular coordinate $\theta$ and $\phi$. It is obvious that the radial and angular tidal forces  depend on the Gauss-Bonnet coupling constant $\alpha$. The change of tidal forces with $\alpha$ is also shown in Fig.\ref{as1}.
As $\alpha\leq0$, we find that the radial tidal force is positive and the angular tidal force is negative for the arbitrary point in the spacetime.
The absolute value of each components of tidal force increases with the absolute value of $\alpha$, which means that the tidal effects in 4D-Einstein-Gauss-Bonnet black hole spacetime with negative $\alpha$ are stronger than that in the Schwarzschild black hole spacetime \cite{s1,s2,s3,Schwarzschild TF}.
However, in the case with positive $\alpha$,
we find that the radial tidal force disappears at the position with
\begin{equation}\label{radial-0}
r=R_{0}^{r t f}=1.930(\alpha M)^{\frac{1}{3}},
\end{equation}
and the angular tidal force vanishes at
\begin{equation}\label{angular-0}
r=R_{0}^{a t f}=(\alpha M)^{\frac{1}{3}}.
\end{equation}
In the region $r>R_{0}^{r t f}$, we find from Fig.\ref{as1} that the strength of tidal forces decreases with $\alpha$, which means that for the case with positive $\alpha$, the tidal forces in 4D-Einstein-Gauss-Bonnet black hole spacetime become weaker than that in the Schwarzschild case. This is different from that in the case with negative $\alpha$, which could be provide a potential way to detect the sign of Gauss-Bonnet coupling constant $\alpha$ by analyzing tidal force if the spacetime geometry of a celestial body can be described by the 4D-Einstein-Gauss-Bonnet black hole metric \eqref{Metric}. From Fig.\ref{as0}, we also find that both $R_{0}^{r t f}$ and $R_{0}^{a t f}$ increase with the coupling constant $\alpha$.  Moreover,
the surface $r=R_{0}^{r t f}$ lies inside the black hole event horizon as $\alpha<0.603$, and outside the event horizon as $\alpha>0.603$, while the surface $r=R_{0}^{a t f}$ always lies between the event horizon and the Cauchy horizon. This means that in the region outside the event horizon, the angular tidal force is always negative, while the radial tidal force would change its sign as $\alpha=0.603$.
For the positive $\alpha$, the radial tidal force has a peak value as
\begin{eqnarray}\label{radial-max}
r=R_{\max }^{r t f}=2.7320(\alpha M)^{\frac{1}{3}}.
\end{eqnarray}
$R_{\max }^{r t f}$ increases with $\alpha$, but the peak value of the radial tidal force can be approximated as $0.037285/\alpha$, which is a decreasing function of $\alpha$.
Similarly,  for the angular tidal force, there is a nadir value  as
\begin{eqnarray}\label{angu-min}
r=R_{\min }^{a t f}=1.5874(\alpha M)^{\frac{1}{3}},
\end{eqnarray}
and the corresponding nadir value is $-0.0669873/\alpha$. For the case with $\alpha\leq0$, the radial and angular tidal forces are the monotonous function of radial coordinate $r$ and there is no peak or  nadir value for the tidal forces. Moreover, we find that the tidal forces are divergent in the limit $r\rightarrow 0$ for the positive $\alpha$, but for the negative $\alpha$ case, they diverge as $r\rightarrow 2(|\alpha|M)^{\frac{1}{3}}$, which does not appear in Reissner-Nordstr\"{o}m black hole spacetime \cite{RNBH1}. Actually, it is understandable because for the case with negative $\alpha$, the surface $r= 2(|\alpha|M)^{\frac{1}{3}}$ becomes a singular surface at where the curvature invariants $R$, $R_{\mu\nu}R^{\mu\nu}$ and $R_{\mu\nu\rho\sigma}R^{\mu\nu\rho\sigma}$ are divergent \cite{gbcurv}.

\section{ Numerical solution of the geodesic deviation equation in the 4D-Einstein-Gauss-Bonnet black hole spacetime }
In this section we solve the geodesic deviation equations (\ref{RTF}) and (\ref{ATF}) to obtain the deviation vector
$\eta^{\hat{\alpha}}$ as functions of $r$, and then analyze the deformation of a neutral body infalling radially under tidal forces in a 4D-Einstein-Gauss-Bonnet black hole spacetime. By using $dr/d\tau=-\sqrt{E^{2}-f(r)}$, one can find that the geodesic deviation equations (\ref{RTF}) and (\ref{ATF}) can be rewritten as \cite{Schwarzschild TF,RNBH,RNBH1}
\begin{equation}\label{radial-dv}
\bigg(E^{2}-f\bigg) \frac{d^2\eta^{\hat{1}}}{dr^2}-\frac{f^{\prime}}{2} \frac{d\eta^{\hat{1}}}{dr}+\frac{f^{\prime \prime}}{2} \eta^{\hat{1}}=0,
\end{equation}
\begin{eqnarray}\label{angular-dv}
\left(E^{2}-f\right)\frac{d^2 \eta^{\hat{i}}}{dr^2}-\frac{f^{\prime}}{2} \frac{d\eta^{\hat{i}}}{dr}+\frac{f^{\prime}}{2 r} \eta^{\hat{i}}=0.
\end{eqnarray}
As in the usual static spacetimes \cite{Schwarzschild TF,RNBH,RNBH1}, the general solutions for  these two differential equations can be expressed as
\begin{eqnarray}
&&\eta^{\hat{1}}=\sqrt{E^{2}-f}\bigg[C_1+C_2\int \frac{dr}{(E^{2}-f)^{3/2}}\bigg],\label{so1}\\
&&\eta^{\hat{i}}=r\bigg[C_3+C_4\int \frac{dr}{r^2\sqrt{E^{2}-f}}\bigg].\label{so2}
\end{eqnarray}
Here the coefficients $C_i$, $i=1,2,3,4$, are the constants of integration, which can be determined by some initial conditions.
We here focus on two types of initial conditions representing dust of particles starting at
the region outside the event horizon ( i.e., $r=b > r_+$) \cite{Schwarzschild TF,RNBH,RNBH1}. The first type of initial conditions $\text { ICI }$ is \cite{Schwarzschild TF,RNBH,RNBH1}
\begin{eqnarray}\label{IC1}
\eta^{\hat{\alpha}}(b)= 1,\quad\quad \dot{{\eta}}^{\hat{\alpha}}(b)=0,
\end{eqnarray}
which corresponds to a body constituted of dust released without initial velocity at $r=b$.  The second kind of initial condition ($\text { ICII }$) is \cite{Schwarzschild TF,RNBH,RNBH1}
\begin{eqnarray}\label{IC2}
\eta^{\hat{\alpha}}(b)= 0, \quad\quad \dot{{\eta}}^{\hat{\alpha}}(b)=1,
\end{eqnarray}
which corresponds to a body constituted of dust ``exploding" from a point $r=b$. With these initial conditions, we can study the radial and angular components of the geodesic deviation vector in the 4D-Einstein-Gauss-Bonnet black hole spacetime.
\begin{figure}
\begin{center}
\includegraphics[width=5.7cm ]{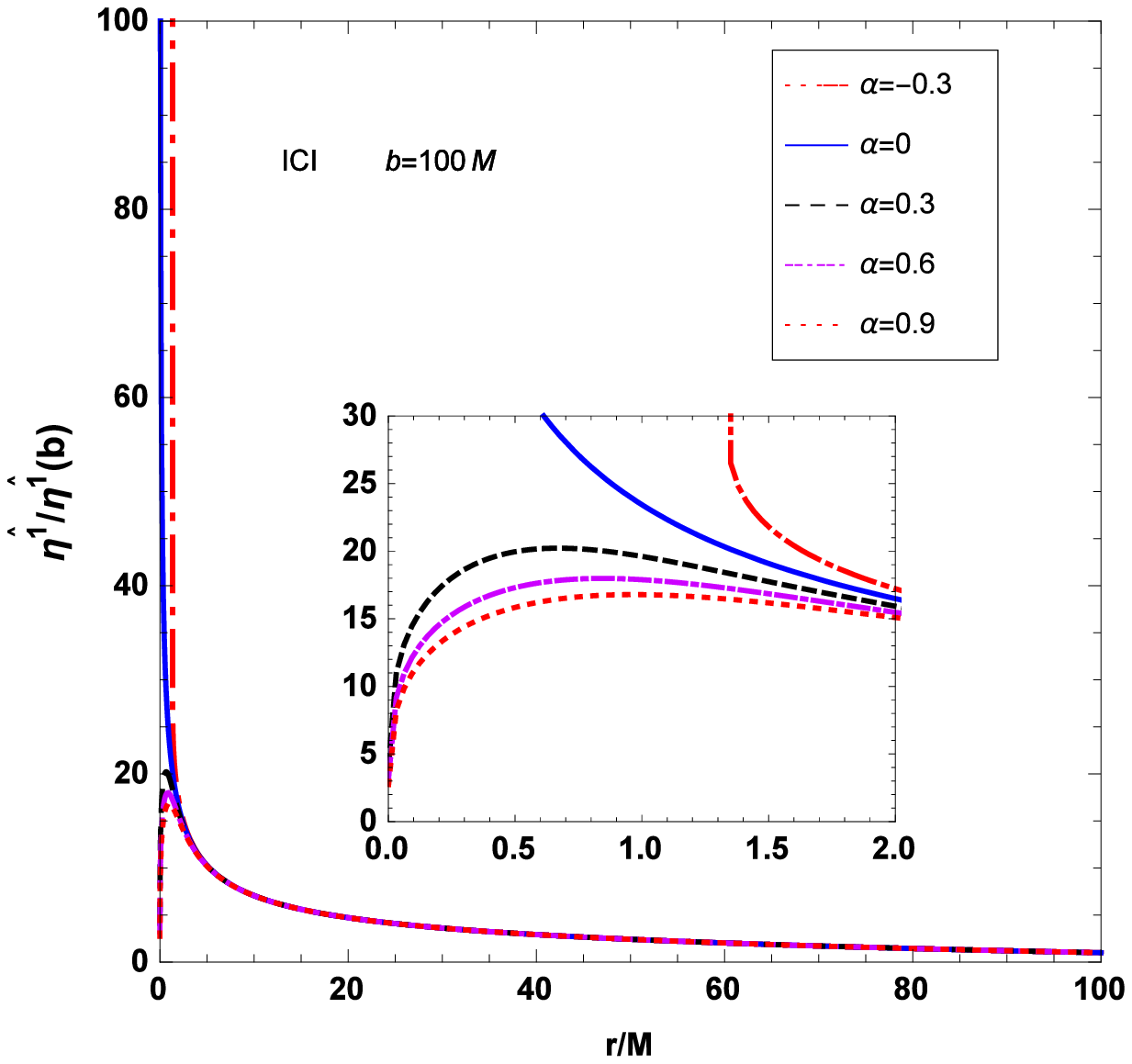}\quad \includegraphics[width=6cm ]{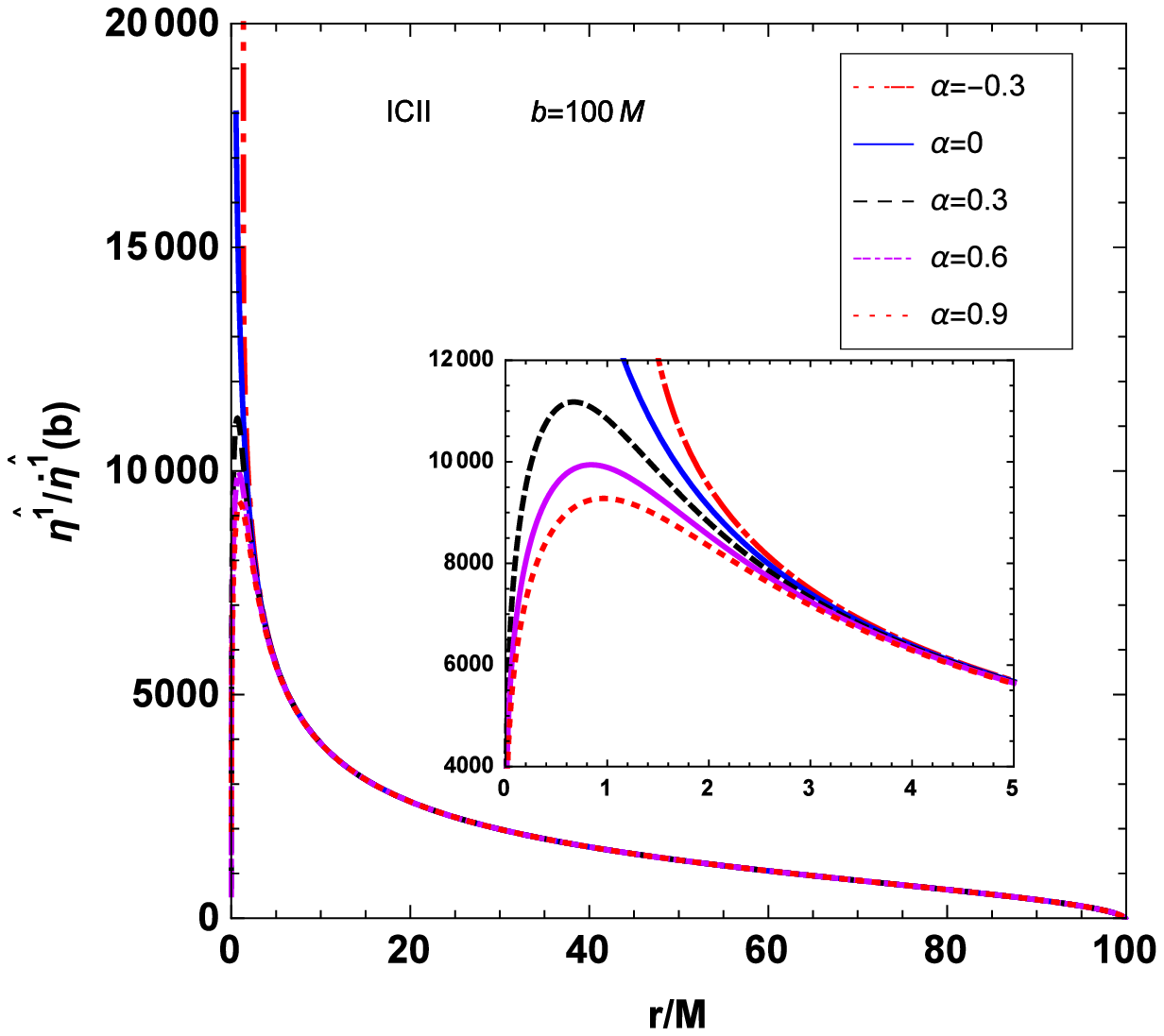}\\
\includegraphics[width=5.7cm ]{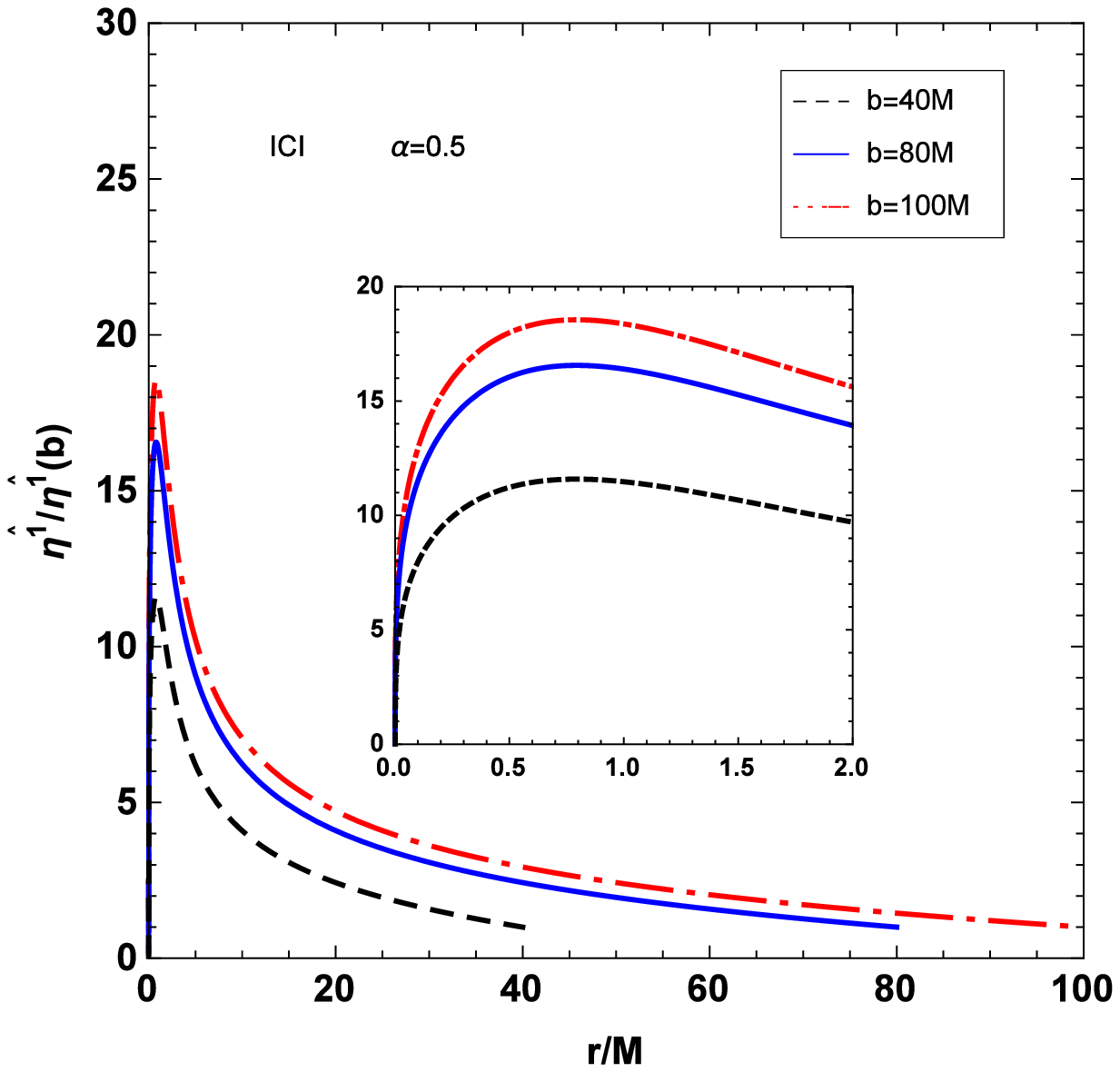} \quad \includegraphics[width=6cm ]{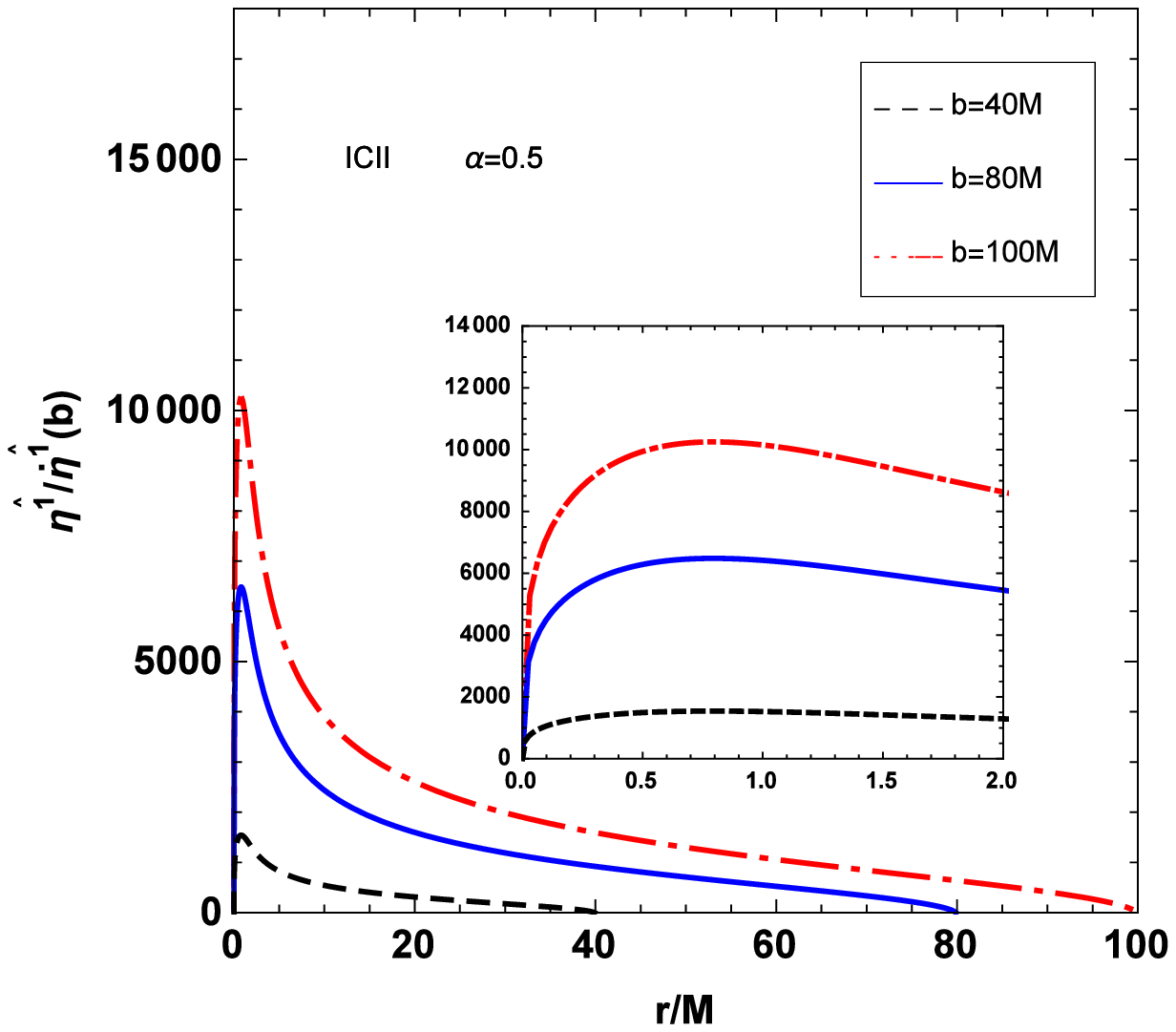}
\caption{ Radial components $\eta ^{\hat{ 1 } }$ of the deviation vector for different the Gauss-Bonnet coupling constant $\alpha$ and the initial radial positions $b$. The left and right panels are related to the initial conditions $\text { ICI }$ and $\text { ICII }$, respectively.  }
\label{as3}
\end{center}
\end{figure}
\begin{figure}
\includegraphics[width=5.0cm ]{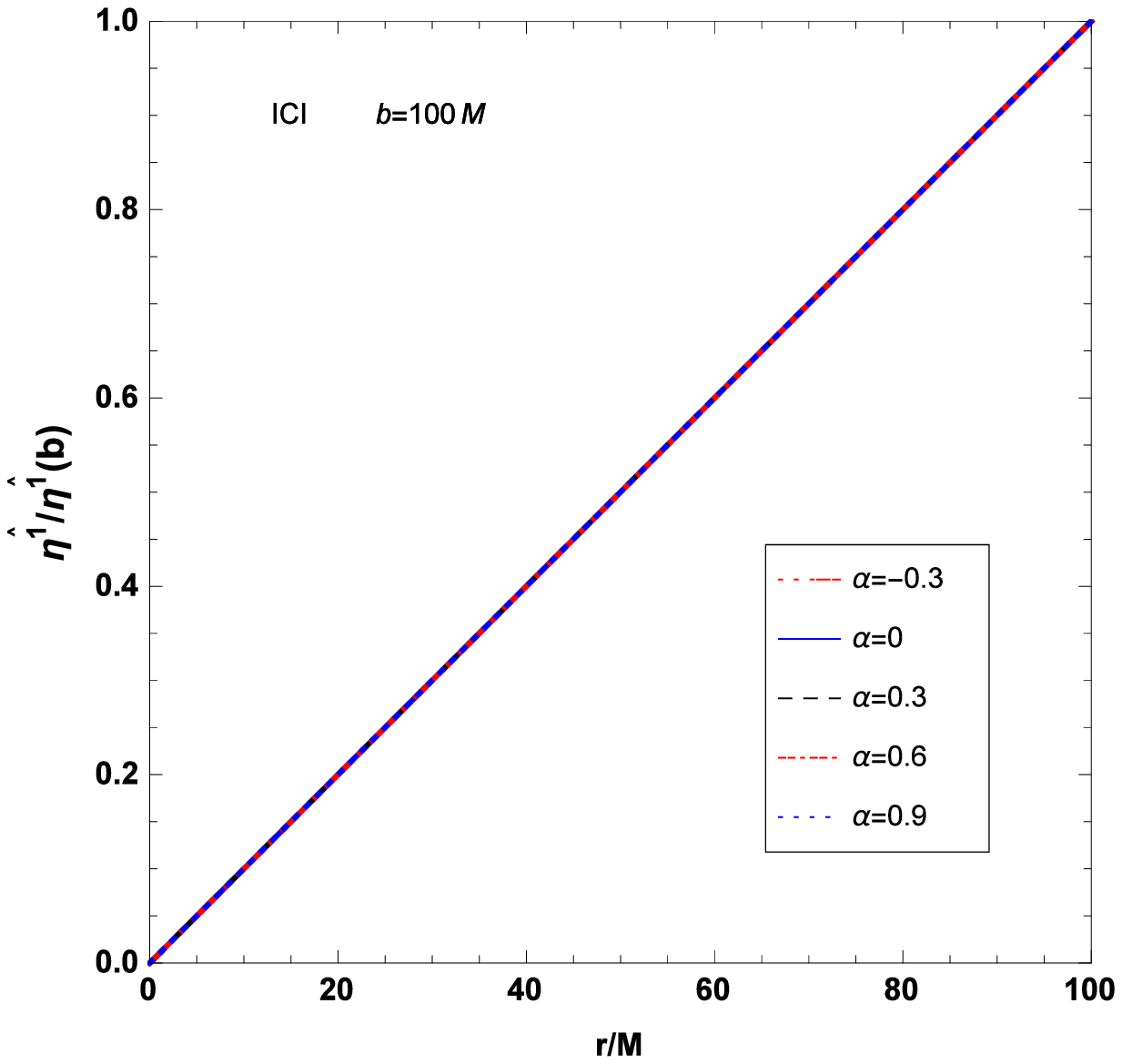} \quad \includegraphics[width=5.7cm ]{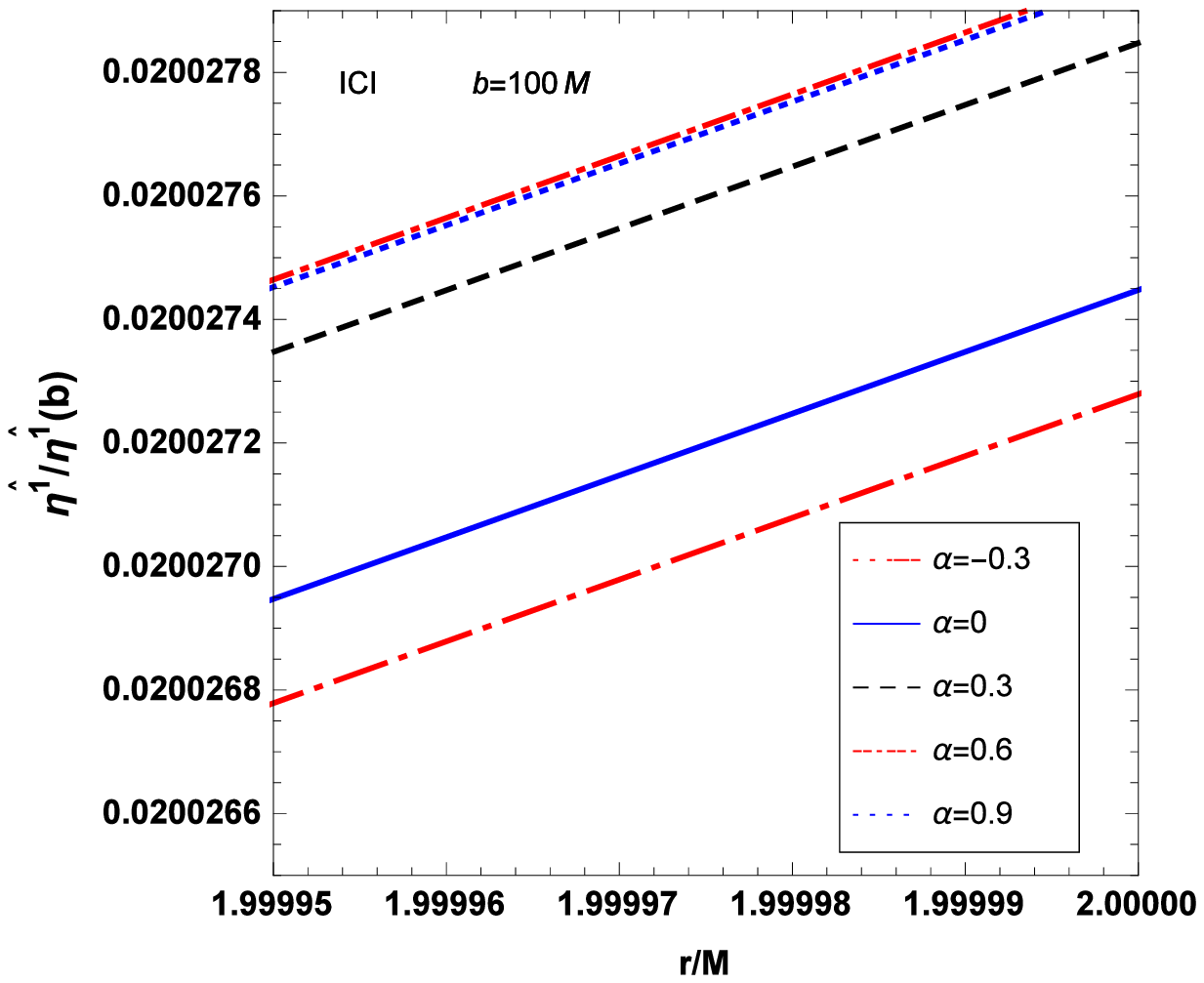}\quad
\includegraphics[width=5.0cm ]{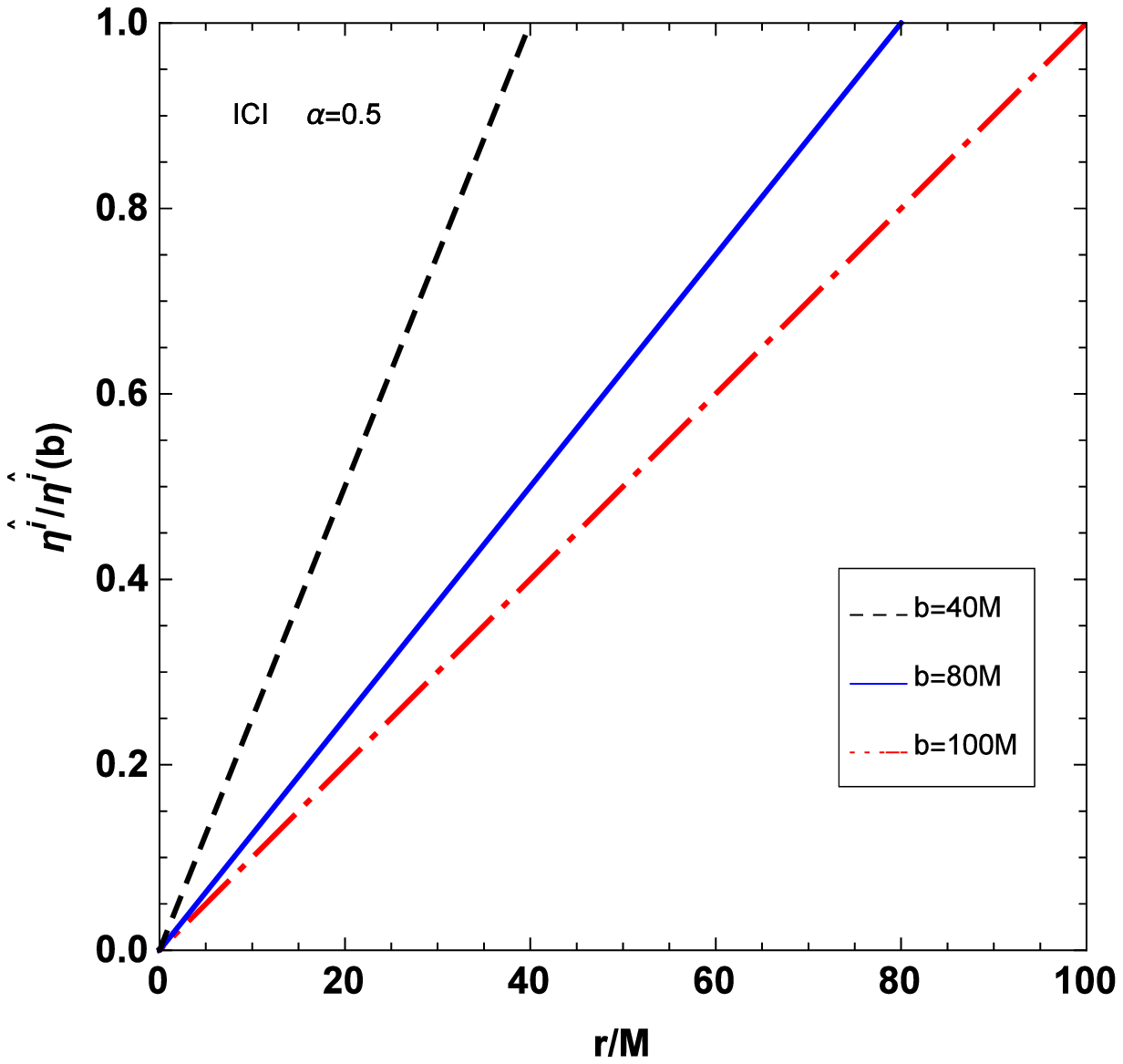} \\
\includegraphics[width=5.2cm ]{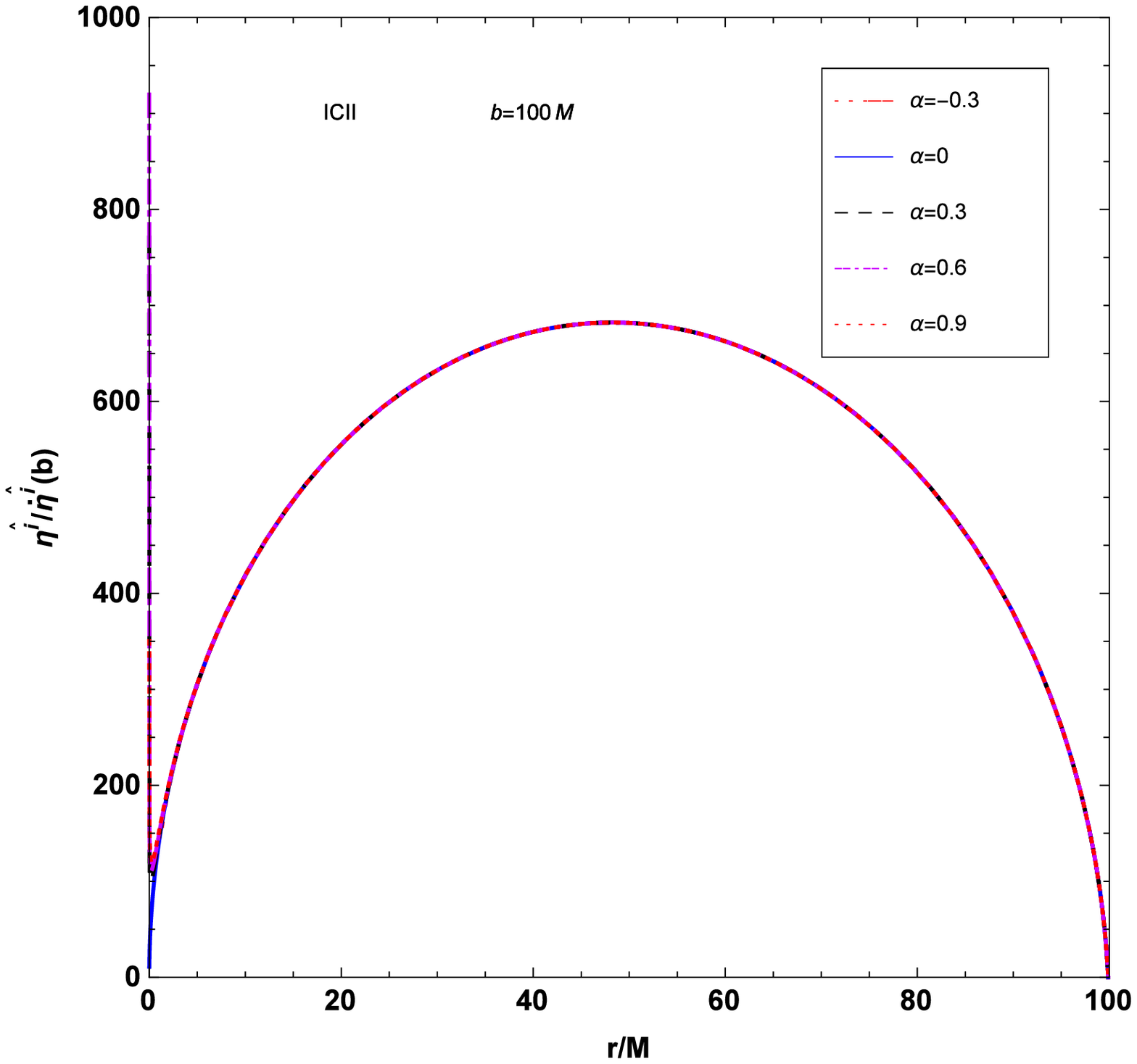} \quad \includegraphics[width=5.12cm ]{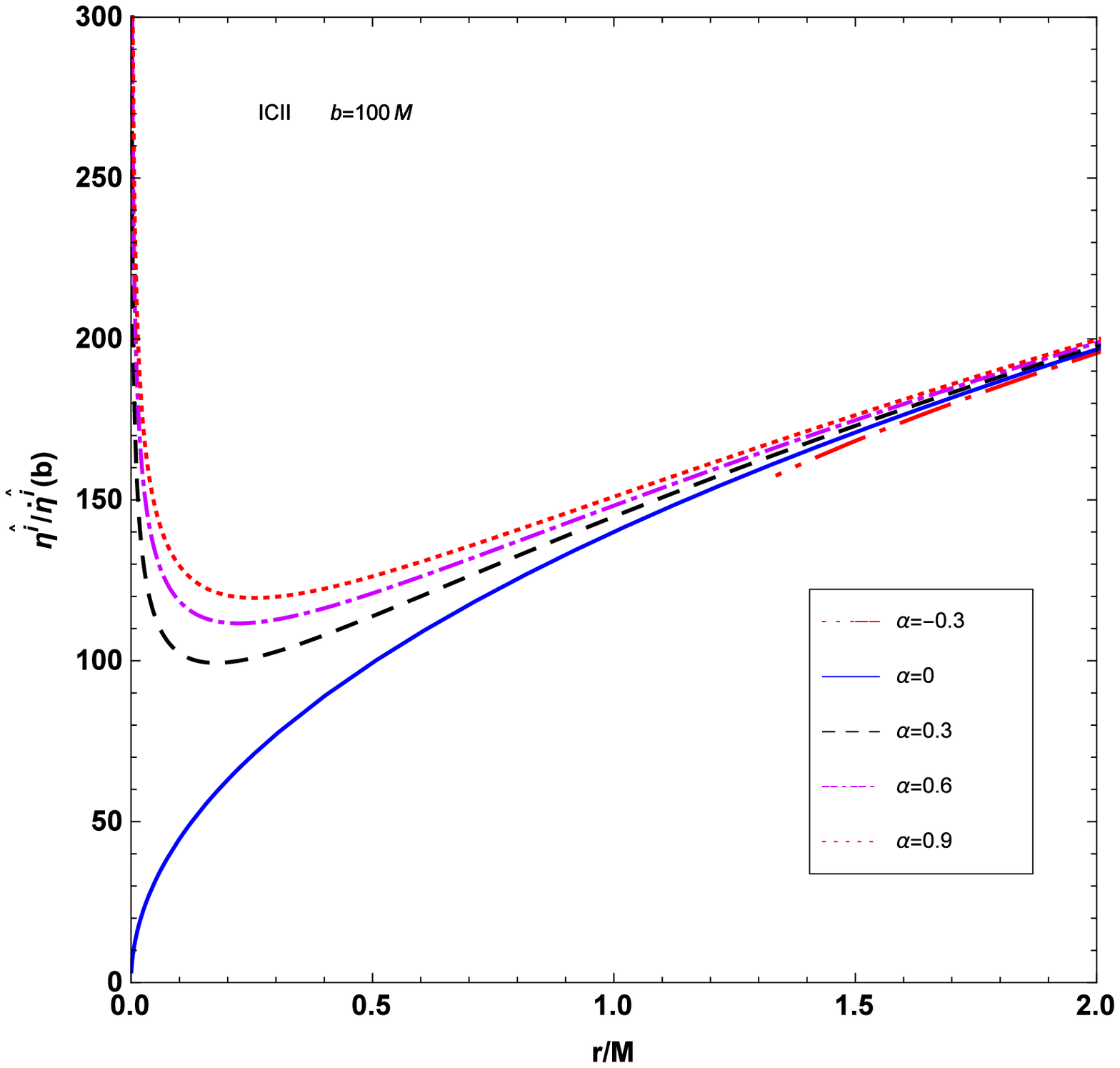}\quad
\includegraphics[width=5.12cm ]{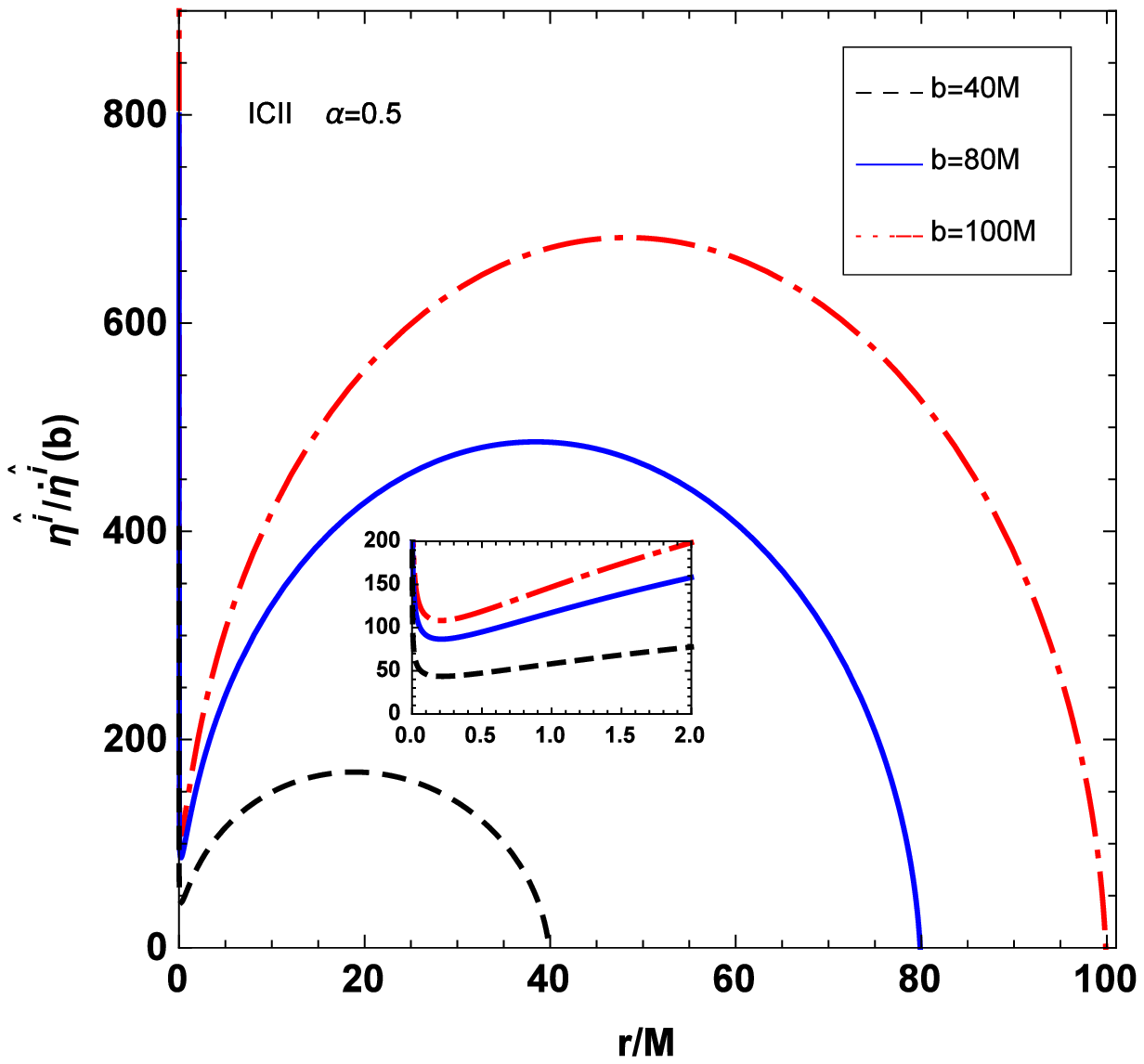}
\caption{ Angular components $\eta ^{\hat{ i } }$ of the deviation vector for different the Gauss-Bonnet coupling constant $\alpha$ and the initial radial positions $b$. The left and right panels are related to the initial conditions $\text { ICI }$ and $\text { ICII }$, respectively. The middle panel in each row is the corresponding partial enlarged detail of the deviation vector in the left panel. }
\label{as4}
\end{figure}

In Fig.\ref{as3}, we find that the radial component of geodesic deviation vector $\eta^{\hat{1}}$  within two conditions $\text { ICI }$ and $\text { ICII }$ increases as the body falls from $r=b$ for the negative $\alpha$, but increases firstly and then decrease for the positive one. Moreover, the change of the radial component $\eta^{\hat{1}}$ with $\alpha$ shows that the radial tidal effects is stronger than that in Schwarzschild spacetime for the  negative $\alpha$, but is weaker
for the positive $\alpha$, which is consistent with the behavior of the radial tidal force in the previous discussion. With the increase of $b$, the radial component of geodesic deviation vector $\eta^{\hat{1}}$ increase under two kinds of initial conditions $\text { ICI }$ and $\text { ICII }$, which means that the radial effects of test body becomes stronger for the initial position with larger $b$.
In Fig. \ref{as4}, we present the change of the angular component of geodesic deviation vector $\eta^{\hat{i}}$  under two conditions $\text { ICI }$ and $\text { ICII }$.  With in the initial condition $\text { ICI }$, $\eta^{\hat{i}}$ for different $\alpha$ decreases as the body falls from $r=b$, which is different from that for the radial component $\eta^{\hat{1}}$. Moreover, it is stronger than that in Schwarzschild spacetime for the positive $\alpha$, but is weaker for the negative $\alpha$, which is the opposite to that for the radial component $\eta^{\hat{1}}$. For the positive $\alpha$, we also find that $\eta^{\hat{i}}$ increases firstly and then decreases with $\alpha$, which is different from the change of the radial tidal force with the Gauss-Bonnet coupling constant $\alpha$. This means that the change of the angular component of geodesic deviation vector is determined not only by the tidal force, but also by the initial conditions. Moreover,  $\eta^{\hat{i}}$ increases with $b$ for the fixed $\alpha$. For the initial condition $\text { ICII }$, as the body falls from $r=b$, the component $\eta^{\hat{i}}$ first increases with the decreasing of $r$ and approaches a maximum as $r$ decreases down to $r=b/2$, which is similar to that in other spacetimes. With the further decreasing of $r$, the angular component $\eta^{\hat{i}}$ decreases. Finally, as the body tends further to the singularity $r=0$, $\eta^{\hat{i}}$ deceases monotonously for the negative $\alpha$, but first decreases and then increases for the  positive  $\alpha$. Moreover,  the angular component $\eta^{\hat{i}}$ for fixed $\alpha$ increases with $b$, which differs from that under the initial condition $\text {ICI }$.

\section{Summary}

 We have investigated the tidal forces and geodesic deviation motion in the 4D-Einstein-Gauss-Bonnet black hole spacetime. Our results show that the tidal force and geodesic deviation motion depend sharply on the sign of the Gauss-Bonnet coupling constant $\alpha$. For the negative $\alpha$, the black hole owns a single black hole horizon and the radial component of the tidal force is always positive and the angular component is negative. The absolute value of tidal force increases as the test body falls from $r=b$. These behaviors are similar to those in the Schwarzschild spacetime with a single horizon. Moreover, the absolute value of tidal force is an increasing function of $\alpha$, which means that the tidal effect is stronger than that in the Schwarzschild spacetime. In the cases with positive $\alpha$, as the test body falls from $r=b$ to $r=0$, the radial component of tidal force first increases and then decrease, the change of the angular component is on the contrary. These properties are similar to those in Reissner-Nordstr\"{o}m black hole with two horizons. The presence of the positive $\alpha$ make the strength of tidal force weaken, which means that the tidal effects in the 4D-Einstein-Gauss-Bonnet black hole spacetime with $\alpha$ is less than that in Schwarzschild case. Moreover, there exists some special surfaces at where the radial or angular component of tidal force disappears. We also find that the tidal forces are divergent in the limit $r\rightarrow 0$ for the positive $\alpha$, but for the negative $\alpha$ case, they diverge as $r\rightarrow 2(|\alpha|M)^{\frac{1}{3}}$, which does not appear in Reissner-Nordstr\"{o}m black hole spacetime.

 We also present the change of geodesic deviation vector in the 4D-Einstein-Gauss-Bonnet black hole spacetime under two conditions $\text { ICI }$ and $\text { ICII }$. The dependence of geodesic deviation vector on $b$ is similar to that in other black hole spacetimes. The change of the radial component $\eta^{\hat{1}}$ with $\alpha$ shows that the radial tidal effects is stronger than that in Schwarzschild spacetime for the  negative $\alpha$, but is weaker for the positive $\alpha$. However, the change of the angular component with $\alpha$ is the opposite to that for the radial component $\eta^{\hat{1}}$. In other words,  it is stronger for the positive $\alpha$, but is weaker for the negative $\alpha$. This means that the  tidal effects on the geodesic deviation vector are determined not only by the tidal force, but also by the initial conditions. These behaviors of tidal forces and  geodesic deviation vector could help us to understand tidal effects and the Gauss-Bonnet gravity in four dimensional spacetime.

\section{\bf Acknowledgments}
We would like to thank Prof. Yen Chin Ong for his useful discussions. This work was  supported by the National Natural Science
Foundation of China under Grant No.11875026, 11875025, 12035005 and 2020YFC2201403.

\end{document}